\documentclass[aps,prr,twocolumn,superscriptaddress,longbibliography,nobalancelastpage]{revtex4-1}

\usepackage{graphicx}
\usepackage{amsmath}
\usepackage{amsthm}
\usepackage{color}
\usepackage[normalem]{ulem}
\usepackage[utf8]{inputenc}
\usepackage{hyperref}

\usepackage{enumitem} 
\usepackage{float} 
\usepackage{wrapfig}
\usepackage{amsfonts}

\usepackage[finnish,swedish,english]{babel}
\usepackage{amsmath,amssymb,amsthm} 

\usepackage{graphicx}
\usepackage{dcolumn}
\usepackage{bbm}
\usepackage{siunitx} 

\usepackage{float} 

\newcommand{\ket}[1]{\left| #1\right\rangle}
\newcommand{\bra}[1]{\left\langle #1\right|}
\newcommand{\Nph}{N_{\mathrm{ph}}}
\newcommand{\Nv}{N_{\mathrm{v}}}
\newcommand{\Nm}{N_{\mathrm{m}}}

\begin{document}
	\title{Mode switching dynamics in organic polariton lasing}
	
	\author{Antti J. Moilanen}
	\email{amoilanen@ethz.ch}
	\affiliation{Department of Applied Physics, Aalto University School of Science, FI-00076 Aalto, Finland}
	\affiliation{Photonics Laboratory, ETH Z\"urich, CH-8093 Z\"urich, Switzerland}
	
	\author{Krist\'in B. Arnard\'ottir}
	\affiliation{DTU Fotonik, Technical University of Denmark, 2800 Kongens Lyngby, Denmark}
	\affiliation{NanoPhoton-Center for Nanophotonics, Technical University of Denmark, 2800 Kongens Lyngby, Denmark}
	\affiliation{SUPA, School of Physics and Astronomy, University of St Andrews, KY16 9SS St Andrews, United Kingdom}
	
	\author{Jonathan Keeling}
	\affiliation{SUPA, School of Physics and Astronomy, University of St Andrews, KY16 9SS St Andrews, United Kingdom}
	
	\author{P\"aivi T\"orm\"a}
	\email{paivi.torma@aalto.fi}
	\affiliation{Department of Applied Physics, Aalto University School of Science, FI-00076 Aalto, Finland}

	\date{\today}
	
	\begin{abstract}
		We study the dynamics of multimode polariton lasing in organic microcavities by using a second-order cumulant equation approach. By inspecting the time evolution of the photon mode occupations, we show that if multiple lasing peaks are observed in time-integrated mode occupations, the reason can be either bi-modal lasing or temporal switching between several modes. The former takes place within a narrow range of parameters while the latter occurs more widely. We find that the origin of the temporal switching is different in the weak- and strong-coupling regimes. At weak coupling slope efficiency is the determining factor, while for strong coupling it is changes in the eigenmodes and gain spectrum upon pumping. This difference is revealed by investigating the photoluminescence and momentum-resolved gain spectra. Our results underscore the importance of understanding the time evolution of the populations when characterizing the lasing behavior of a multimode polariton system, and show how these features differ between weak and strong coupling.
		
	\end{abstract}
	
	\maketitle
	
	\section{Introduction}
	
	In a system where lasing occurs,
	switching between lasing modes can occur, as a function of parameters (e.g. excitation power) or as a function of time. Mode switching is known to occur in semiconductor lasers that support several densely spaced longitudinal modes. Multiple lasing peaks in the time-integrated luminescence spectrum, as well as temporal switching between the modes, have been observed in many different contexts~\cite{furfaro_mode-switching_2004,pedaci_experimental_2005,shore2005unlocking,gordon_multimode_2008,wright_spatiotemporal_2017}. 
	The origin and nature of such mode switching differs between systems. For example, the mode switching in edge-emitting semiconductor lasers and vertical cavity surface emitting lasers (VCSELs) has been attributed to spatial hole burning~\cite{herre_mode_1989,lenstra_rate-equation_2014,valle_dynamics_1995,law_effects_1997} and to four-wave mixing~\cite{yacomotti_dynamics_2004}. Understanding the nature of mode switching in a given system, and how to control it, is an important step toward enabling applications.
	
	Exciton-polaritons---polaritons for brevity---are the hybrid of a photon and an exciton under strong light-matter coupling.
	Strong coupling can be engineered by creating optical microcavities filled with an active layer of emitter material. The active layer can take several forms, including inorganic semiconductors, two-dimensional materials such as transition metal dichalcogenides, or---the focus of this paper---organic molecules.
	Polariton lasing (or polariton condensation) has been studied extensively during the past two decades~\cite{Deng2010a,Keeling2011,keeling_boseeinstein_2020}. 
	
	While the majority of the experimental and theoretical works have focused on single-mode polariton lasing, multimode operation has been reported as well. Switching of macroscopic polariton population between different spatial modes or momentum states has been observed as a function of pump power~\cite{sun_stable_2018,kusudo_stochastic_2013,kim_coherent_2016,maragkou_spontaneous_2010,sawicki_polariton_2021}.
	The switching between higher- and lower-energy polariton modes has been attributed to polariton-polariton scattering, which is stimulated by Bose enhancement at higher pump powers, driving the relaxation of polaritons to the lowest-energy mode~\cite{doan05:prb,Doan2006a,maragkou_spontaneous_2010,sawicki_polariton_2021}. Typically mode switching has been observed only for specific detuning conditions. In coupled wells or lattices of coupled polariton condensates, competition between different modes has been studied experimentally~\cite{lai_coherent_2007,Kim2011,jacqmin_direct_2014,amo_exciton-polaritons_2016,scafirimuto_tunable_2021,boulier_microcavity_2020} and numerically by using the generalized Gross-Pitaevskii equations~\cite{eastham_mode_2008,wouters_synchronized_2008}. Recently, it was noted that simultaneous polariton lasing can occur in a vertical lasing mode and a horizontal guided mode in planar microcavities, and that such phenomena could have taken place in several previous experiments where polariton lasing in the vertical lasing mode
	has been reported with ZnO-based microcavities~\cite{jamadi_competition_2019}. 
	Mode competition dynamics has also recently been explored in InGaAs based cavities~\cite{Topfer2020}.
	In closely related experiments on photon Bose-Einstein condensates (BECs)---with weak light-matter coupling in dye-filled microcavities---multimode condensation has been observed experimentally and studied by second-order cumulant equations~\cite{vlaho_non-equilibrium_2021,vlaho_controlled_2019,marelic_spatiotemporal_2016,kurtscheid_thermally_2019}. Similarly, for strongly-coupled plasmonic BECs, multiple peaks near the ground state have been observed~\cite{Vakevainen2020}.

	In our previous work~\cite{arnardottir_multimode_2020}, we introduced a new model for multimode polariton lasing in organic microcavities and computed steady-state lasing phase diagrams as a function of pump strength and detuning of the photon modes from the exciton. We showed that at certain pump strengths multiple lasing peaks may be observed in the steady-state photon mode occupations. 
	Our second-order cumulant model goes beyond mean-field descriptions, capturing the effects of non-lasing modes and fluctuations on the system's behavior. Retaining such fluctuations is important when considering switching: the (initially small) populations of non-lasing modes act as the seed for these modes to become macroscopically populated at later times. In contrast, for mean-field approaches, explicit seed or noise terms would be required.
	
	In this paper, we investigate the dynamics of mode switching in multimode polariton lasing, and analyze the gain spectrum in more depth than in~\cite{arnardottir_multimode_2020}, including momentum-resolved gain spectra. While the model we consider is specific to organic molecules, the concepts we discuss have relevance more widely. We find that in the case where several lasing peaks are observed in the time-integrated mode occupations, the modes are not necessarily lasing \textit{simultaneously}, but the lasing mode may switch (deterministically) as a function of time. Furthermore, even when only a single lasing peak is observed, switching between different modes can occur before the system reaches a steady state.
	We study mode switching in both weak and strong coupling regimes. 
	
	Mode switching occurs in general because the mode which lases first does not fully cross-saturate the gain for other modes. This raises a question of what determines which other modes become competitive, and ultimately take over.
	At weak coupling, the choice of which mode is selected can be explained by a standard argument~\cite{siegman1986lasers} considering the different thresholds and slope efficiencies (gradient of photon population vs pump) of each mode:  When several modes are above the threshold, the mode with highest gain will eventually suppress the others. At strong coupling, the situation becomes more complicated because of energy shifts. The polariton energies show a blue shift as a function of pump strength, due to effective interactions that arise from saturation of the gain medium. Importantly, we also show that the peak gain shifts towards higher energies as the pump strength increases. This shift can be attributed to dynamics of the vibrational degrees of freedom of the molecules. The mode switching at strong coupling occurs as a result of competition between both shifts in the gain and the polariton energies. Our model considers spatially homogeneous systems, thus our results suggest that mode switching does not necessarily require redistribution of real space intensity. 
	
	The rest of the paper is organized as follows. In Section~\ref{sec:sysmod}, we present the system and the model. In Section~\ref{sec:mmlasing}, we study the temporal characteristics of multimode lasing. In Section~\ref{sec:gain}, we present the gain and photoluminescence spectra as a function of pump strength. In Section~\ref{sec:pulsed} we consider pulsed excitation, which is relevant for typical experiments on organic exciton-polariton systems. Finally, in Section~\ref{sec:disc} we discuss the interpretation and implications of the results. Appendices provide technical details on the cumulant equations and results for time-integrated populations.
	
	\section{System and model}
	\label{sec:sysmod}
	
	\subsection{Density matrix equation of motion}
	We consider a planar microcavity that hosts multiple photon modes and an active layer of emitters. We model the emitters as $\Nm$ vibrationally dressed two-level systems placed homogeneously in the cavity plane. The system is thus described by the Tavis--Cummings--Holstein Hamiltonian in the rotating wave approximation (RWA)~\cite{Cwik2014,Strashko2018,arnardottir_multimode_2020}:
	\begin{align}
		H =& \sum_{\mathbf k} \omega_{\mathbf k} a^\dagger_{\mathbf k} a_{\mathbf k} + \sum_{n,{\mathbf k}} \left(g_{n,{\mathbf k}}a_{\mathbf k} \sigma_n^+ + g_{n,{\mathbf k}}^\ast a_{\mathbf k}^\dagger \sigma_n^-\right) \nonumber\\
		& + \sum_n \left[\frac{\varepsilon}{2}\sigma^z_n + \omega_v\left(b_n^\dagger b_n + \sqrt{S}\left(b_n^\dagger + b_n\right)\sigma_n^z\right)\right].
		\label{eq:Hamiltonian}
	\end{align}
	Here $\varepsilon$ is the transition energy of the molecule and Pauli matrices $\sigma_n^{z,+,-}$ correspond to the electronic states of molecule $n$. The operator $b_n^\dagger$ creates a vibrational excitation of energy $\omega_v$ on molecule $n$ while $a_{\mathbf k}^\dagger$ creates a photon with energy $\omega_{\mathbf k}$ in the cavity mode with wave vector ${\mathbf k}$. 
	
\begin{figure}
		\centering
		\includegraphics[width=1.0\columnwidth]{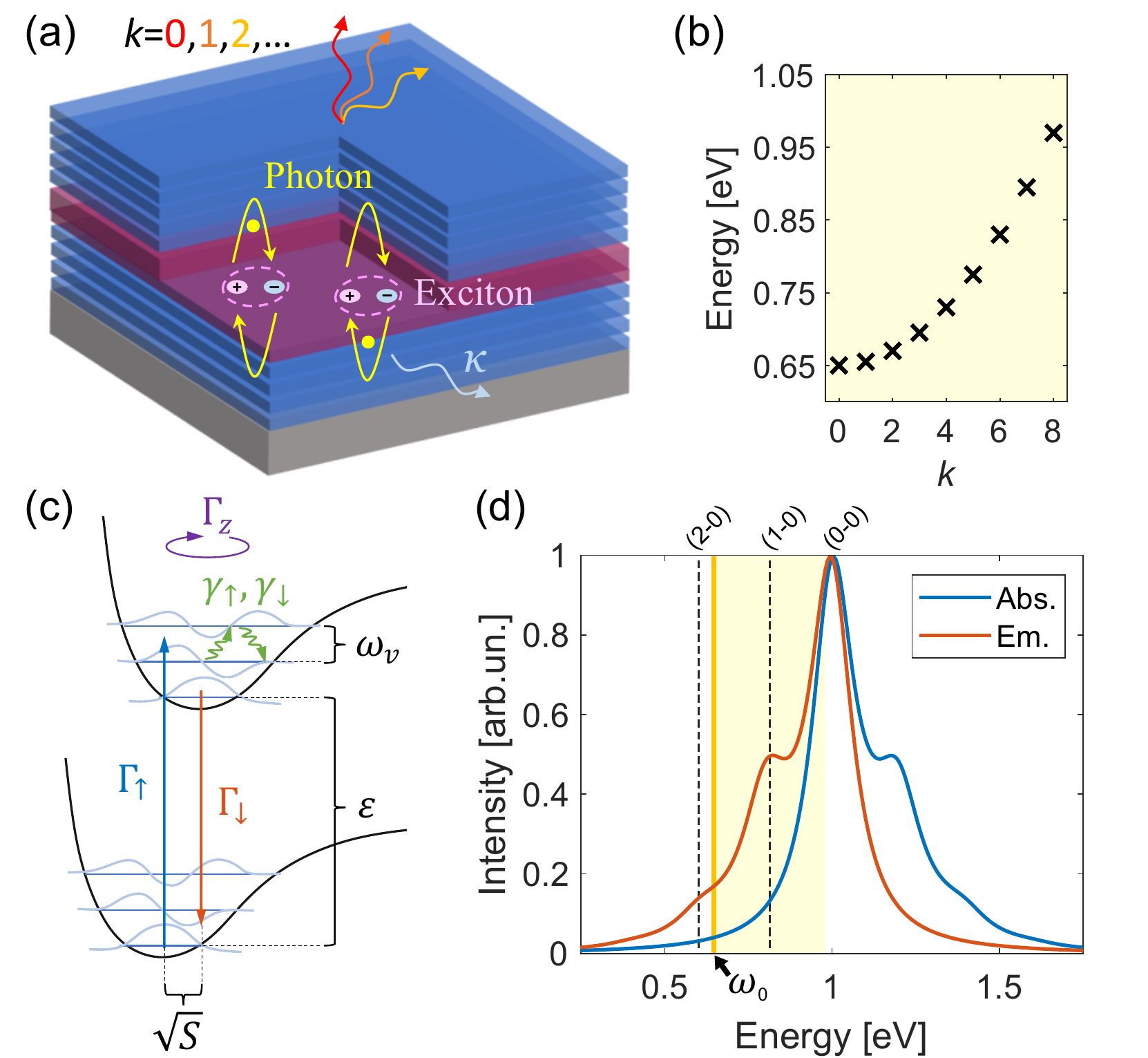}
		\caption{Schematics of the system and model. (a) A planar microcavity with multiple photon modes and a layer of emitter material. The cavity modes have in-plane momentum $k$ and they decay by rate $\kappa$. The photons in the cavity modes can strongly couple to the excitons, hybridizing to form exciton-polaritons. (b) Discretized dispersion relation of the uncoupled cavity modes. (c) Two-level system with both ground and excited state dressed by $\Nv$ vibrational states. The bare transition energy between the two electronic states  is denoted by $\varepsilon$, the vibrational frequency by $\omega_v$, and the Huang-Rhys factor by $S$. The rates of the drive and dissipation processes involved are the external pumping $\Gamma_\uparrow$, spontaneous decay $\Gamma_\downarrow$, dephasing $\Gamma_z$, and thermal excitation and dissipation of the vibrational modes $\gamma_{\uparrow,\downarrow}$. (d) Absorption and emission spectra of the uncoupled emitters. The yellow shaded area corresponds to the energy range which the uncoupled photon modes cover, starting from $\omega_0$ (0.65~eV). 
			The zero-phonon-line (0-0) and the emission shoulders related to the two first vibrational transitions (1-0), (2-0) are labeled.}
		\label{fig:Fig1}
	\end{figure}

	The photon energies are given by $\omega_{\mathbf k} = \omega_0 + {E_\rho}{(K_x^2+K_y^2)}/{N_\textrm{m}}$, where $K_{x,y}$ are integers labelling a discrete set of momentum vectors in the two-dimensional plane of the cavity. The energy scale $E_\rho =  \pi^2 \rho_\text{2D}{\hbar^2}/{2 m}$ is defined in terms of the molecular density $\rho_{2D}$ and effective photon mass $m$. The total number of photon modes is truncated to $\Nph$. The light-matter coupling is given by the coupling constant $g_{n,{\mathbf k}}=e^{-i \mathbf{k} \cdot \mathbf{r}_n} \Omega_R/\sqrt{\Nm}$, where $\mathbf{r}_n$ is the position of the $n$th molecule. 
	These plane-wave cavity modes correspond to considering periodic boundary conditions.
	Coupling of the electronic levels to vibrational modes results in broadening of emission and absorption spectra and in the occurrence of a Stokes shift, which is parametrized  by the Huang-Rhys factor $S$~\cite{jong_resolving_2015}.
	In later discussions it will be useful to use the notation $(n-m)$ to denote transitions between the $n$th vibrational ground state and the $m$th vibrational excited state; see Fig.~\ref{fig:Fig1}(c,d).
	The total number of vibrational states on each molecule will be truncated to $\Nv$.
	
	The incoherent processes that occur can be described by the Lindblad formalism. The equation of motion for the density matrix of the system reads
	\begin{multline}
		\label{eq:Lindblad}
		\partial_t \rho =  -i[H,\rho] + \sum_\mathbf{k} \kappa \mathcal{L}[a_\mathbf{k}]
		+ \sum_n\big(\Gamma_\uparrow\mathcal{L}[\sigma^+_n]
		+ \Gamma_\downarrow\mathcal{L}[\sigma^-_n]\\
		+ \Gamma_z \mathcal{L}[\sigma^z_n]
		+ \gamma_\uparrow \mathcal{L}[b^\dagger_n + \sqrt{S}\sigma^z_n]
		+ \gamma_\downarrow \mathcal{L}[b_n + \sqrt{S}\sigma^z_n] \big),
	\end{multline}
	with $\mathcal{L}[X] = X\rho X^\dagger - \frac{1}{2}(X^\dagger X \rho + \rho X^\dagger X)$. The equations account for cavity loss at rate $\kappa$, dephasing of the excitons $\Gamma_z$, and incoherent pumping and decay of excitons with rates $\Gamma_\uparrow$ and $\Gamma_\downarrow$, respectively \footnote{We note that compared to our previous work Ref.~\cite{arnardottir_multimode_2020} we have corrected the sign of the $\sqrt{S}\sigma^z_n$ term in Eq.~\eqref{eq:Lindblad}. This does not significantly change the results.}. The last two terms describe thermal relaxation and excitation with rates $\gamma_\downarrow = \gamma_v (n_b+1)$ and $\gamma_\uparrow = \gamma_v n_b $ where $n_b = \left[\text{exp}(\omega_v/k_B T_v) - 1\right]^{-1}$ is the Bose-Einstein distribution at temperature $T_v$. The values of all parameters used for the simulations below are given in Table~\ref{tab:params}.

	\begin{table}
		\centering
		\caption{\textbf{Simulation parameters}.}
		
		\begin{center}
			\begin{tabular}{l | l | l}
				Parameter & Value & Notes \\ \hline\hline
				$\Nm$                & $10^8$        &   \\
				$N_\text{v}$                    & 4             &  
				\\
				$\Nph$               & 9             &  
				\\\hline
				$\varepsilon$ 	        & 1~eV          &  
				\\
				$\omega_0$             & 0.65~eV 	    & 
				$  - 2 \omega_v < \omega_0 -\epsilon < -\omega_v$
				\\ 
				$E_\rho$               &$5\times 10^{5}$~eV & \\
				$\Omega_\text{R}$ & 0.1~eV \& 0.4~eV &   \\
				$\omega_v$  	        & 0.2~eV 	    &   \\
				S                      & 0.10          &   \\\hline
				$\kappa$               & $10^{-4}$~eV  &   $\sim$~(40ps)$^{-1}$ \\
				$\Gamma_\downarrow$  	& $10^{-4}$~eV  &   $\sim$~(40ps)$^{-1}$\\ 
				$\Gamma_z$  	        & 0.03~eV	    & 	$\sim$~(1ps)$^{-1}$\\
				$\gamma_v$  		    & 0.02~eV       &   \\
				$k_BT_v$			    & 0.025~eV 	    &  Room temperature
				\label{tab:params} 
			\end{tabular}
		\end{center}
		
	\end{table}
	
	\subsection{Second-order cumulant equations}
	
	The size of Hilbert space for $\Nm$ molecules, each with $2 \Nv$ internal states, is too large to directly simulate except for small $\Nm$.  
	To make the simulation computationally tractable for realistic system sizes, 
	we proceed from Eq.~\eqref{eq:Lindblad} by using the second-order cumulant expansion. This provides a (closed) set of equations of motion for first- and second-order correlations between molecule and photon operators~\cite{Kirton2017,SanchezBarquilla2020,arnardottir_multimode_2020}.

	In order to make the cumulant expansion, it is helpful to first relabel the molecular degrees of freedom. The two electronic states $(\sigma)$ and the $\Nv$ vibrational states ($b$) are combined into $2\Nv$-level operators, which can be represented by generalized Gell-Mann matrices, $\lambda$ (see Appendix~\ref{app:GGM} for details). The Hamiltonian then becomes:
	\begin{equation}
		\label{eq:Hamiltonian_GM}
		H = \sum_\mathbf{k} \omega_\mathbf{k} a_\mathbf{k}^\dagger a_\mathbf{k} + \sum_n \left[A_i + \sum_\mathbf{k}(B_i a^\dagger_\mathbf{k} e^{-i \mathbf{k} \cdot \mathbf{r}_n} + \textrm{h.c.})\right]\lambda_i^{(n)},
	\end{equation}
	where the vectors $A_i$ and $B_i$ include the terms in Eq.~\eqref{eq:Hamiltonian}. Tensor sums that run over Gell--Mann operator indices $i$ are implicit. In the new basis, the master equation in Eq.~\eqref{eq:Lindblad} takes the form 
	\begin{equation}
		\label{eq:Lindblad_GM}
		\partial_t \rho = -i[H,\rho] +  \sum_{\mathbf{k}} \kappa \mathcal{L}[a_{\mathbf{k}}] +\sum_{\mu,n} \mathcal{L}\left[\gamma_i^{\mu} \lambda^{(n)}_i\right],
	\end{equation}
	where $\mu$ labels the drive and dissipation processes related to molecules: (1) pump, (2) decay, (3) dephasing, and vibrational (4) excitation and (5) decay. 
	
	We next write down equations of motion for first- and second-order correlations of the operators $\lambda^{(n)}_i, a^{}_{\mathbf{k}}, a^{\dagger}_{\mathbf{k}}$. To produce a closed set of equations of motion, all third- or higher-order correlations are split into products of first and second-order correlations, by setting third-order cumulants (e.g.~$\langle ABC\rangle_c$ ) to zero. This allows to decompose the third-order terms as $\langle ABC\rangle = \langle A\rangle \langle BC\rangle +\langle B\rangle \langle AC\rangle +\langle C\rangle \langle AB\rangle$~\cite{gardiner2009stochastic}.
	To further simplify the derivation, we make use of the conservation of number of excitations implied by the RWA and divide the operators $\lambda_i$ into three groups that either increase ($+$), decrease ($-$), or conserve ($z$) the electronic excitations. With this classification, expectation values of operators that do not conserve the number of excitations can be taken zero, such as the first-order terms $\langle a_\mathbf{k} \rangle,\langle a_\mathbf{k}^\dag \rangle, \langle \lambda_{i_-} \rangle,\langle \lambda_{i_+} \rangle$. 
	
	Assuming a spatially homogeneous distribution (or random distribution with a large number) of emitters in the microcavity, we may approximate summation over sites by using $\sum_n e^{i (\mathbf{k}-\mathbf{k}^\prime)\cdot \mathbf{r}_n}=\Nm \delta_{\mathbf{k},\mathbf{k}^\prime}$, allowing the use of momentum conservation. For the number-conserving Gell-Mann operators we then obtain a site-independent form $\ell_i=\langle \lambda_{i_z}^{(n)} \rangle $. For the two other groups of Gell-Mann operators, $\lambda_{i_+}^{(n)}$ and $\lambda_{i_-}^{(n)}$, the first order expectations are zero as they do not conserve the number of excitations. We can write the Fourier components of the second-order correlations as
	$d_{ij}^\mathbf{k}=\sum_{n,m\neq n} e^{i\mathbf{k}\cdot(\mathbf{r}_n-\mathbf{r}_m) }\langle \lambda_{i_+}^{(n)}\lambda_{i_-}^{(m)}\rangle/\Nm^2$ and $c_i^\mathbf{k}=\sum_n e^{i\mathbf{k}\cdot\mathbf{r}_n }\langle a^{}_\mathbf{k} \lambda_{i_+}^{(n)}\rangle/\Nm$. Photon mode occupations are denoted by $n^\mathbf{k}=\langle a^\dagger_{\mathbf{k}} a^{}_{\mathbf{k}} \rangle$. 
	The equations of motion for the non-vanishing correlations are thus:
	\begin{align}
		&\partial_t n^\mathbf{k} = -\kappa n^\mathbf{k} - 2\Nm \mathrm{Im}{B_i c_i^\mathbf{k}}, \nonumber\\
		&\partial_t \ell_i = \xi_{ij} \ell_j +\phi_i - 8 \mathrm{Re}{\beta_{ij} \sum_k c_j^\mathbf{k}}, \nonumber\\
		&\partial_t c_i^\mathbf{k} = \left[X_{ij} -\left(i\omega_\mathbf{k}+\frac{\kappa}{2}\right)\delta_{ij}\right]c_j^\mathbf{k} + 2\beta^\ast_{ij}\ell_j n_\mathbf{k} -iB_j d_{ij}^\mathbf{k} \nonumber\\
		&-\frac{i}{\Nm}\left(\zeta_{ij}\ell_j + \frac{B_i}{2\Nv} \right), \nonumber\\
		&\partial_t d_{ij}^\mathbf{k} = X^\ast_{ip} d^\mathbf{k}_{pj} + X_{jp}d^\mathbf{k}_{ip} +2\ell_p\left(\beta_{ip}\Tilde{c_j}^\mathbf{k\ast} + \beta^\ast_{jp}\Tilde{c_i}^\mathbf{k}\right),
		\label{eq:secondcumulants}
	\end{align}
	where $\Tilde{c_i}^{\mathbf{k}} = c_i^\mathbf{k} -\sum_\mathbf{q} c_i^\mathbf{q}/\Nm$, and the coefficients are given in Appendix~\ref{app:GGM}. Equations~\eqref{eq:secondcumulants} describe the time evolution of the strongly coupled system with multiple photon modes, capturing the effect of fluctuations. Solving these equations yields the time evolution of the photon mode populations, discussed in Section~\ref{sec:mmlasing}.

	\begin{figure*}
		\centering
		\includegraphics[width=1\textwidth]{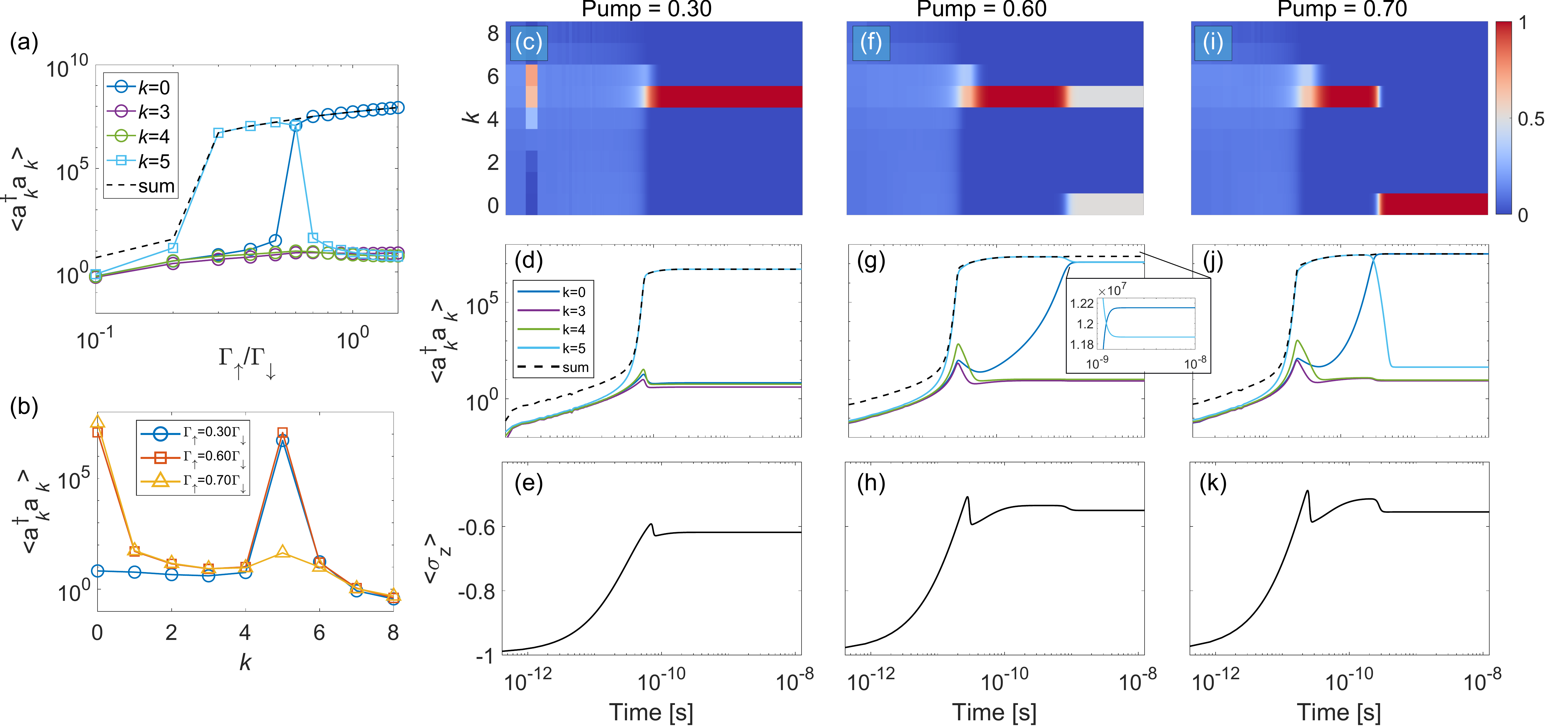}
		\caption{Mode competition for weak coupling, $\Omega_R=0.1$eV. (a) Pump dependence curves and (b) steady-state occupations at different $k$ modes. The mode occupations in (b) are shown for selected pumps. The time-integrated threshold curves are shown in Appendix~\ref{app:timeintegrated}. On the right: (Top and middle row) time evolution of photon population at different modes and (bottom row) population inversion of the two-level system as a function of pump strength. The pump strengths are (c-e) $\Gamma_\uparrow/\Gamma_\downarrow=0.30$, (f-h) $0.60$, and (i-k) $0.70$. All other parameters are provided in Table~\ref{tab:params}.
		}
		\label{fig:FigXX2}
	\end{figure*}

	\subsection{Photoluminescence and gain spectra}
	\label{sec:PL_gain}
	
	When coupling between the cavity modes and the emitters is strong, polariton lasing replaces photon lasing. In this limit, understanding the nature of mode competition becomes more challenging, since both the gain profile and the normal mode energies must be found self consistently, and change with pump strength. In this section we present the numerical approach we will use to extract both quantities, as used in Section~\ref{sec:gain}.
	
	The energies of the polariton modes can be found by computing the photoluminescence spectrum, which is given by
	\begin{equation}
		\label{eq:PL}
		S_{\mathbf{k}}(\nu) = \int_{-\infty}^\infty dt \langle a_{\mathbf{k}}^\dagger (t) a^{}_{\mathbf{k}}(0)\rangle e^{i\nu t}.
	\end{equation}
	Here, we utilize the quantum regression theorem to calculate the two-time correlators~\cite{Breuer2002}. Solving $S_{\mathbf{k}}(\nu)$ requires the steady-state density matrix $\rho_\textrm{ss}$ which is used to construct $\tilde{\rho}_{\mathbf{k}}(0) = a^{}_{\mathbf{k}}\rho_\textrm{ss}$. Time-evolving the effective density matrix $\tilde\rho_{\mathbf{k}}(t) = e^{t\mathcal{L}} \tilde{\rho}_{\mathbf{k}}(0)$ and evaluating $\text{Tr}\big[ a_{\mathbf{k}}^\dagger \tilde{\rho}_{\mathbf{k}}(t) \big]$ gives the coupled differential equations for the two-time correlators~\cite{Scully1997}:
	\begin{align}
		\partial_t \langle a^\dagger_{\mathbf{k}} (t) a^{}_{\mathbf{k}}(0)\rangle &=
		\left(i\omega_k - \frac{\kappa}{2}\right) \langle a^\dagger_{\mathbf{k}} (t) a^{}_{\mathbf{k}}(0)\rangle + i\Nm B_i^\ast c_i^{\mathbf{k}}(t) \nonumber,\\
		\partial_t c_i^{\mathbf{k}}(t) &= \xi_{ij}c_j^{\mathbf{k}}(t) + 2f_{ijp} B_j \ell_p\langle a^\dagger_{\mathbf{k}} (t) a^{}_{\mathbf{k}}(0)\rangle,
		\label{Eq:Fluorb}
	\end{align}
	where $c_i^{\mathbf{k}}(t) = \frac{1}{\Nm}\sum_n e^{i \mathbf{k} \cdot \mathbf{r}_n} \text{Tr}\left[\lambda_{i_+}^{(n)}\tilde\rho_{\mathbf{k}}(t)\right]$ and
	\begin{align*}
		\langle a^\dagger_{\mathbf{k}} (t) a^{}_{\mathbf{k}}(0)\rangle = \text{Tr}\left[a^\dagger_{\mathbf{k}} e^{t\mathcal{L}}a^{}_{\mathbf{k}} \rho_{ss}\right].
	\end{align*}
	In the steady state, $\ell_p$ becomes constant. The linear equations can be written in a matrix form as $\partial_t \mathbf{C_k} = \mathcal{M}\mathbf{C_k}$, where the vector $\mathbf{C_k} = [\langle a^\dagger_{\mathbf{k}} (t) a^{}_{\mathbf{k}}(0)\rangle, \{c_i^{\mathbf{k}}(t)\}_i]$ and the matrix $\mathcal{M}$ are obtained from Eqs.~\eqref{Eq:Fluorb}. The Fourier transform can be written as
	\begin{align}
		S_\mathbf{k}(\nu) &= \int_{-\infty}^{\infty} e^{i\nu t + \mathcal{M}t} \mathbf{C}(0) dt \nonumber\\ 
		&= (i\nu + \mathcal{M})^{-1} \mathbf{C}(0) = \sum_i \frac{\alpha_i}{\mu_i+i\nu} \ket{r_i},
	\end{align}
	where $\mu_i$ is the eigenvalue of $\mathcal{M}$ that corresponds to the right $\ket{r_i}$ and left $\bra{l_i}$ eigenvectors. The coefficient $\alpha_i$ is then given by $\alpha_i=\bra{l_i}\mathbf{C}(0)\rangle/\bra{l_i}r_i\rangle$.
	
	The ``gain'' (i.e., emission minus absorption) can be studied by considering two-time correlations of the molecules:
	\begin{align}
		\label{eq:GainTotal}
		G(\nu) &=
		\int_{-\infty}^\infty dt e^{-i\nu t}  
		\langle[\sigma^{+(n)} (t),\sigma^{-(n)}(0)]\rangle
		\nonumber\\&=
		V_i^+V_j^-\int_{-\infty}^\infty dt e^{-i\nu t}  
		\langle[\lambda_i^{(n)} (t),\lambda_j^{(n)}(0)]\rangle,
	\end{align}
	where $V_i^\pm = \frac{1}{2}\text{Tr}(\sigma^\pm\lambda_i)$. In the weak-coupling picture, this has a simple interpretation as the frequency-resolved gain the photon modes see; at strong coupling the picture is more complicated, but this nonetheless allows one to probe the molecular dynamics separately from the photons.

	By also considering the time-dependent correlations between different molecules, one can further compute the momentum-resolved gain spectrum:
	\begin{align}
		\label{eq:GainResolved}
		G_{\mathbf{k}}(\nu) =& V_i^+V_j^-\int_{-\infty}^\infty dt e^{-i\nu t} \\ \nonumber &\sum_{n,m} e^{i\mathbf{k}\cdot(\mathbf{r}_n-\mathbf{r}_m)}\langle[\lambda_i^{(n)} (t), \lambda_j^{(m)}(0)]\rangle.
	\end{align}
	This quantity identifies the different gain seen by each cavity mode, and so helps explain when mode switching occurs. Note that we can recover the form of $G(\nu)$ in Eq.~\eqref{eq:GainTotal} by summing $G_{\mathbf{k}}(\nu)$ over all $\mathbf{k}$.
	
	\section{Multimode lasing dynamics with continuous pump}
	\label{sec:mmlasing}
	
	\subsection{Weak-coupling regime}
	
	Using the methods described in the previous section, we now explore the dynamics of mode switching, starting with the weak-coupling regime, for which we set  $\Omega_R=0.1$eV.
	As shown in Fig.~\ref{fig:FigXX2}(a,b) (similar results also seen in our previous work~\cite{arnardottir_multimode_2020}), for the parameters we use the lasing mode switches between $k=5$ and $k=0$ as a function of pump strength. Upon increasing the pump strength, the mode $k=5$, which is closest to resonance with the $(1-0)$ transition, starts lasing first. When the pump is further increased, the $k=0$ starts to lase, and the lasing at $k=5$ is suppressed. As noted in Ref.~\cite{arnardottir_multimode_2020}, the selection of which mode wins mode competition in the weak coupling regime can be attributed to the different threshold and slope efficiency---i.e., gradient of photon population vs pump---of each mode. The modes near the vibrational shoulder ($k=4,5,6$) have the lowest threshold, but the mode near $k=0$ has the highest slope efficiency (better visible in linear-scale threshold curves, see~\cite{arnardottir_multimode_2020}). Thus at low pumping (low inversion), lasing occurs at $k=5$, whereas at high pumping (high inversion), the mode $k=0$ with the higher slope efficiency wins. 
	
	The time evolution of the system, as shown in Fig.~\ref{fig:FigXX2}(c-k), reveals that mode switching occurs not only as a function of pump strength, but also in time. Here, the term `switching' corresponds to the time instant when the population of one lasing mode surpasses the population of another mode. Fig.~\ref{fig:FigXX2}(c,f,i) show for illustrative purpose the data shown in Fig.~\ref{fig:FigXX2}(d,g,j), which presents the population of selected $k$ modes. Fig.~\ref{fig:FigXX2}(e,h,k) shows the population inversion as a function of time.
	
	At a low pump strength, in Fig.~\ref{fig:FigXX2}(c-d), lasing starts and remains at $k=5$ until steady state is reached. At higher pump strengths, as shown in Fig.~\ref{fig:FigXX2}(f-g) and Fig.~\ref{fig:FigXX2}(i-j), switching to $k=0$ mode occurs. At a high pump strength, Fig.~\ref{fig:FigXX2}(i-j), lasing at $k=0$ completely suppresses the lasing at $k=5$ after the switch. Only in a very narrow range of pump strengths, as illustrated in Fig.~\ref{fig:FigXX2}(f-g), does the system show bi-modal lasing. In that case, the $k=0$ mode slowly grows until it marginally prevails, and both modes sustain a macroscopic population at steady state. The dashed lines in Fig.~\ref{fig:FigXX2}(d,g,j) show the photon population summed over all $k$ modes. It is interesting to note that the sum over all $k$ modes shows only a single threshold; whenever switching takes place the photon population is re-distributed between different modes. One may note also that the population inversion (as shown in Fig.~\ref{fig:FigXX2}(e,h,k)) appears to reach a stationary value before switching occurs. After the switch, the inversion does however show a small kink.
	
	\begin{figure}
		\centering
		\includegraphics[width=0.75\columnwidth]{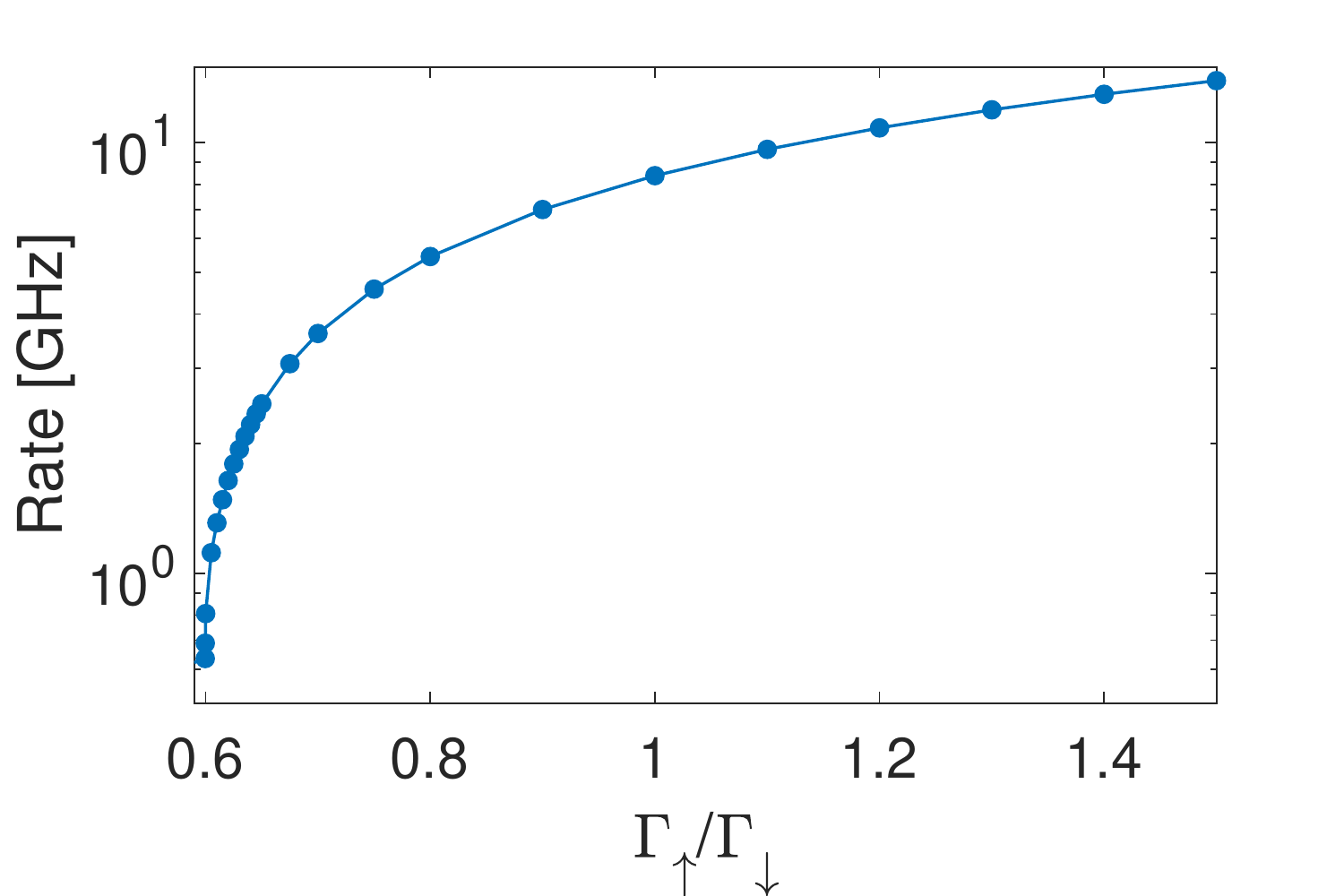}
		\caption{Metastable decay rate of the $k=5$ mode as a function of pump strength, in the vicinity of the transition point. The decay rate is the inverse of the time at which the lasing mode switch occurs, i.e., when the population of mode $k=0$ exceeds that of mode $k=5$. Parameters as in Fig.~\ref{fig:FigXX2}.
		}
		\label{fig:FigSXX3}
	\end{figure}

	\begin{figure*}
		\centering
		\includegraphics[width=1\textwidth]{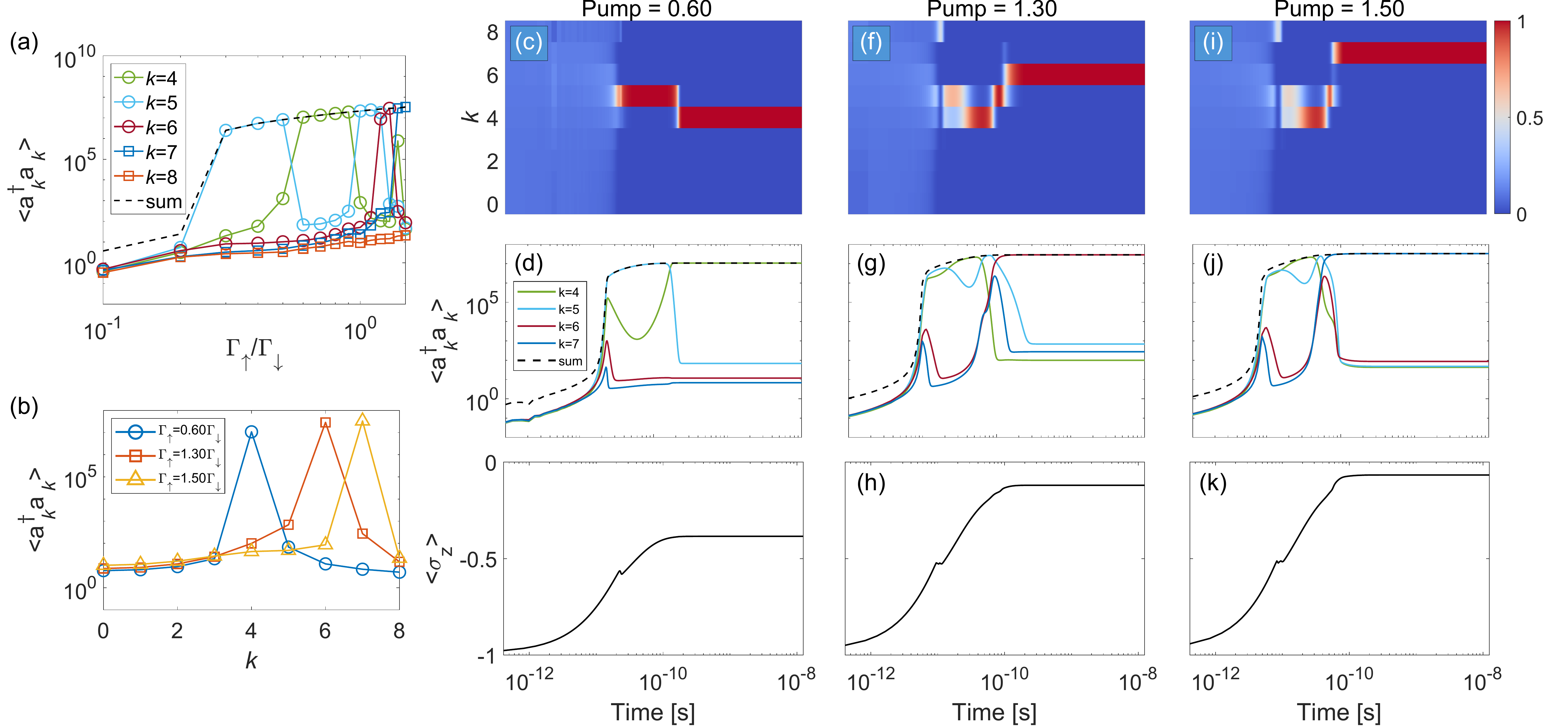}
		\caption{Mode competition for strong coupling, $\Omega_R=0.4$eV. (a) Pump dependence curves and (b) steady-state occupations at different $k$ modes. The mode occupations in (b) are shown for selected pumps. The time-integrated threshold curves are shown in Appendix~\ref{app:timeintegrated}. (Top and middle row) time evolution of photon population at different modes and (bottom row) population inversion of the two-level system as a function of pump strength. The pump strengths are (c-e) $\Gamma_\uparrow/\Gamma_\downarrow=0.60$, (f-h) $1.30$, and (i-k) $1.50$.  All other parameters are provided in Table~\ref{tab:params}.
		}
		\label{fig:FigXX4}
	\end{figure*}
	
	In Figure~\ref{fig:FigSXX3}, we show the rate at which the metastable $k=5$ state is replaced by $k=0$. We see that the rate vanishes as one approaches the transition point, i.e., the pump strength where bi-modal lasing is observed ($\sim~0.6\Gamma_\downarrow$, Fig.~\ref{fig:FigXX2}(f-h)).

	\subsection{Strong coupling regime}
	
	At strong coupling ($\Omega_R=0.4$eV) we observe shifting of the lasing mode towards higher $k$ on increasing the pump strength. As shown by the threshold curves and spectra in Fig.~\ref{fig:FigXX4}(a-b), lasing is first triggered in the $k=5$ mode but switches to $k=7$ at high pump strengths. In a narrow regime of pump strengths, the $k=6$ mode has the highest occupation.
	
	The time evolution of the mode occupations for strong coupling is presented in Fig.~\ref{fig:FigXX4}(c-k). At a low pump strength, in Fig.~\ref{fig:FigXX4}(c-d), lasing starts at $k=5$ and switches to a lower $k=4$ mode. At higher pump strengths, in Fig.~\ref{fig:FigXX4}(f-g) and Fig.~\ref{fig:FigXX4}(i-j), the lasing mode switches to higher $k$ modes. One notable distinction from weak coupling is seen in the time evolution of the population inversion. As seen in Fig.~\ref{fig:FigXX4}(h,k), the population inversion continues to grow until the (final) mode switching occurs. 
	
	Both in the strong and weak coupling regimes, we have referred to the lasing modes with the absolute value of the momentum vector, $k=|\mathbf{k}|$. Within our model, as we have no source of disorder, all modes with a given $|\mathbf{k}|$ are degenerate. In real experiments, the disorder that will be present is
	likely to break the perfect degeneracy, leading to a particular angular pattern being selected for the lasing mode.

	\section{Gain spectrum}
	\label{sec:gain}
	
	To better understand why mode switching occurs in response to pumping, we next discuss how gain and photoluminescence spectra change as a function of pump power. To avoid the subtleties associated with time-dependent spectra~\cite{Eberly77}, we only consider the steady-state spectra.

	\subsection{Weak coupling regime}

	\begin{figure}
		\centering
		\includegraphics[width=1\columnwidth]{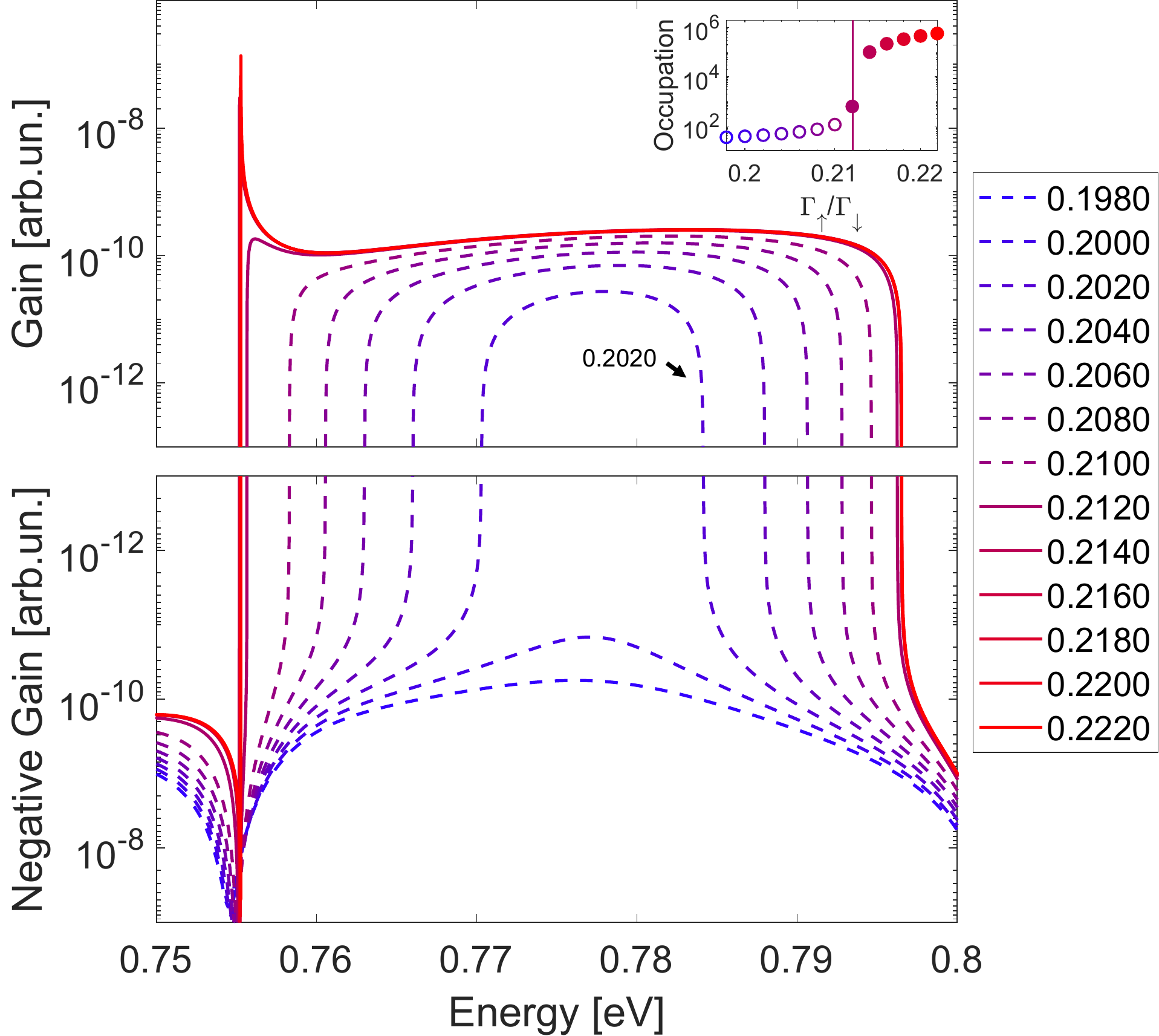}
		\caption{Momentum-integrated gain spectra $G(\nu)$ near the lasing threshold for weak coupling, $\Omega_R=0.1$eV. The top panel shows $G(\nu)$, while the bottom panel shows $-G(\nu)$, both on a logarithmic scale. The pump strengths are color coded as in the legend and the inset which shows the corresponding part of the threshold curve. Solid lines in the panels (and filled circles in the inset) are results above the lasing threshold. Threshold is around 0.21~$\Gamma_{\uparrow}/\Gamma_{\downarrow}$, marked by the vertical line in the inset. All other parameters as in Table~\ref{tab:params}.
		}
		\label{fig:FigXXgain_pump_wc_dense}
	\end{figure} 
	
	\begin{figure}
		\centering
		\includegraphics[width=1\columnwidth]{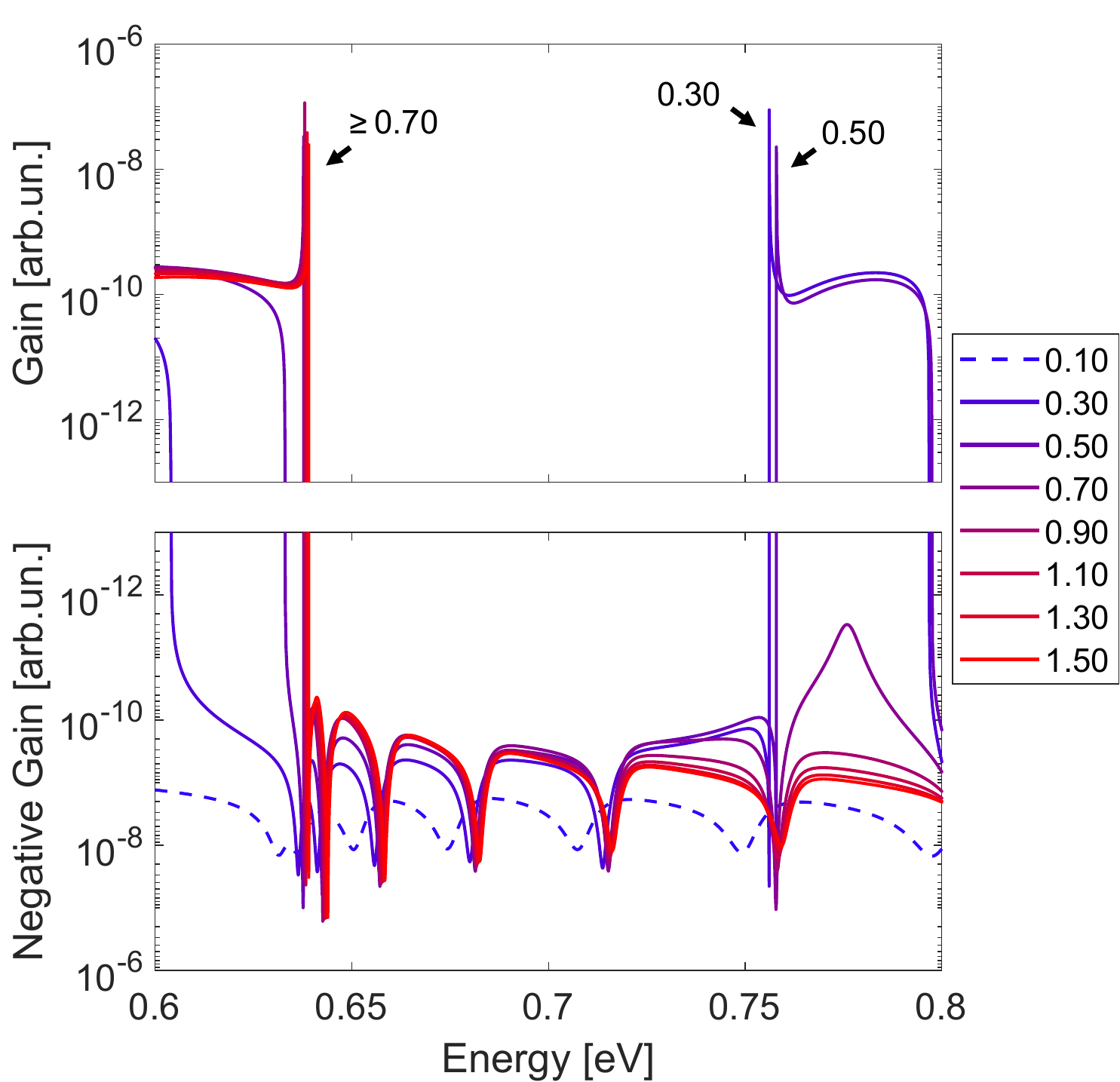}
		\caption{Momentum-integrated gain spectra $G(\nu)$ as a function of pump strength for the weak coupling case $\Omega_R=0.1$eV. This figure shows an expanded range of energies and pump strengths compared to Fig.~\ref{fig:FigXXgain_pump_wc_dense}.
		}
		\label{fig:FigXXgain_pump_wc_sparse}
	\end{figure}

	To introduce the properties of the gain spectrum, we start by considering the total (momentum-integrated) spectrum $G(\nu)$ of the molecules defined in Eq.~\eqref{eq:GainTotal}.  Figure.~\ref{fig:FigXXgain_pump_wc_dense} shows $G(\nu)$ for weak coupling, at a narrow range of pump strengths (see legend) near the initial lasing threshold ($\sim$~0.21~$\Gamma_{\uparrow}/\Gamma_{\downarrow}$). Since $G(\nu)$ can be either positive (net gain from the molecules) or negative (net absorption), we show both $G(\nu)$ and $-G(\nu)$ on logarithmic scales.
	Below threshold (dashed lines) $G(\nu)$ is negative at all frequencies. As the pump strength increases a region of positive net gain develops. Positive gain from the molecules does not immediately lead to lasing, as it is not sufficient to overcome the cavity losses. When the gain at the frequency of the $k=5$ cavity mode (around 0.75~eV) becomes sufficiently high to exceed cavity loss, lasing occurs.
	Figure~\ref{fig:FigXXgain_pump_wc_sparse} then shows how the gain spectrum continues to develop, considering a wider range of frequencies (covering all relevant cavity modes) and a wider range of pump strengths. From this figure, one sees that
	between pump strengths 0.60 and 0.70~$\Gamma_{\uparrow}/\Gamma_{\downarrow}$ the peak of the gain switches to a lower energy mode at 0.63-0.64~eV.  
	
\begin{figure*}
		\centering
		\includegraphics[width=0.9\textwidth]{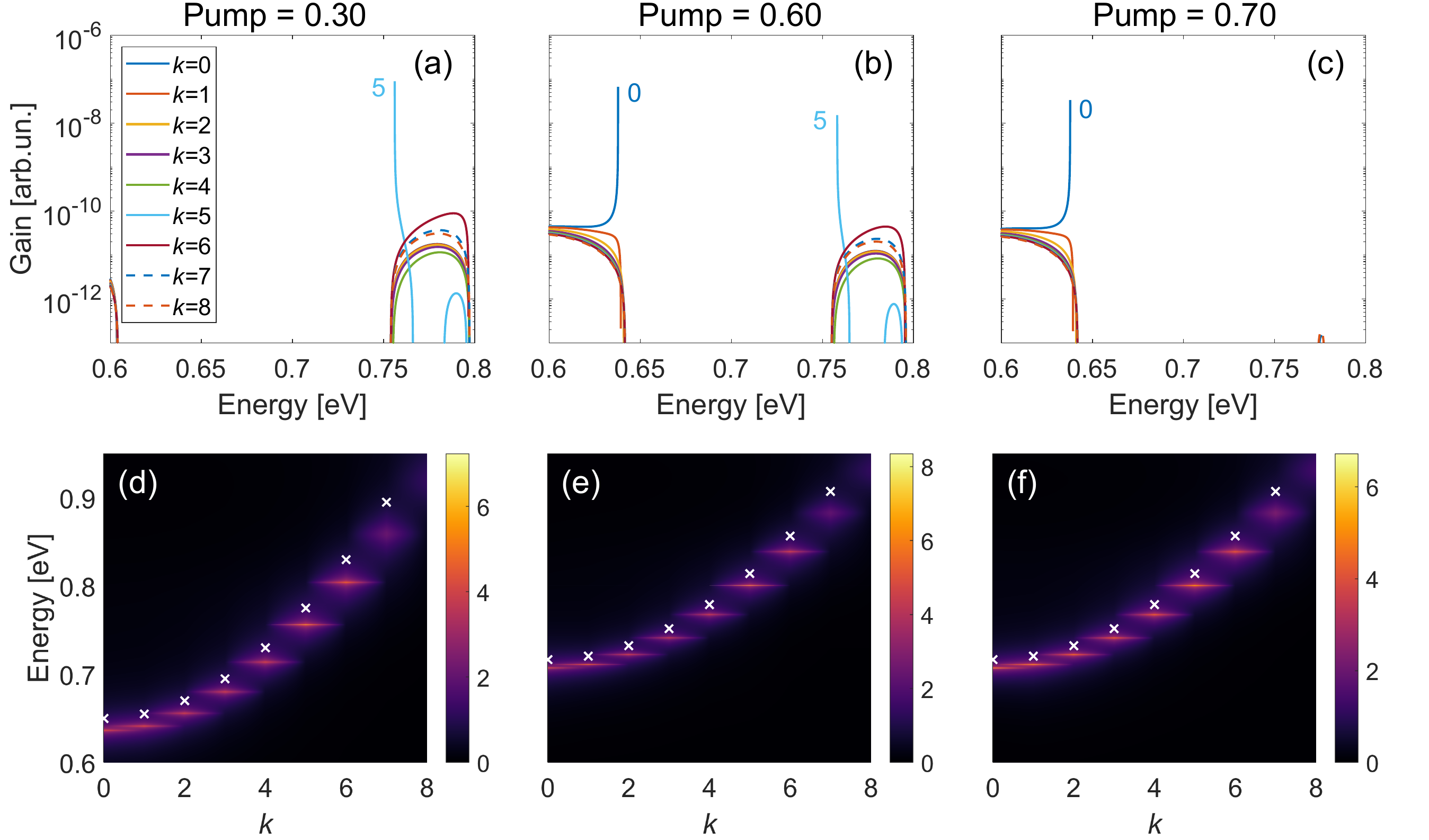}
		\caption{(a-c) Momentum-resolved gain spectra $G_{\mathbf{k}}(\nu)$ and (d-f) photoluminescence spectra $S_{\mathbf{k}}(\nu)$ for selected pump strengths for the weak coupling case, $\Omega_R=0.1$eV. In (a-c) the numbers next to the peaks label the $k$ of the mode(s) that are lasing in the steady state. In (d-f) the white crosses mark the uncoupled cavity modes. The pump strengths are (a,d) $0.30~\Gamma_\uparrow/\Gamma_\downarrow$, (b,e) $0.60~\Gamma_\uparrow/\Gamma_\downarrow$, and (c,f) $0.70~\Gamma_\uparrow/\Gamma_\downarrow$; all other parameters as in Table~\ref{tab:params}. }
		\label{fig:FigXXpl_gain_pump_wc_kdependent}
	\end{figure*}

	We can further study how the gain affects different cavity modes by considering $G_{\mathbf{k}}(\nu)$. 
	The momentum-resolved gain spectra are shown in Fig.~\ref{fig:FigXXpl_gain_pump_wc_kdependent}(a-c) for the three selected pumps that were studied in Section~\ref{sec:mmlasing} (Fig.~\ref{fig:FigXX2}). Here only the positive gain values are plotted. The results confirm that the peak of the gain shifts from the mode $k=5$ to $k=0$ as the pump is increased, and there is a narrow region of pump strengths where the gain is similar for the two modes. We calculated also the photoluminescence spectrum $S_{\mathbf{k}}(\nu)$ to see how the mode energies change upon increasing the pump strength. As shown in Fig.~\ref{fig:FigXXpl_gain_pump_wc_kdependent}(d-f), under weak coupling the modes are only slightly shifted from the bare photon dispersion, and so in this case the mode locations remain practically unchanged as the pump is increased.
	
	\subsection{Strong coupling regime}

	In the strongly coupled system one should now consider lasing arising from scattering into the polariton modes.  As discussed elsewhere~\cite{mazza2013microscopic,Strashko2018}, one can still consider a gain spectrum for feeding into the polariton mode.  This gain spectrum corresponds to the reservoir of dark excitonic states.  
	At strong coupling, the gain spectrum shows more complex changes as a function of pump strength.
	Figure~\ref{fig:FigXXgain_pump_sc_sparse} shows the evolution of total gain $G(\nu)$ with pump strength. As before, below threshold (dashed line) the gain is negative, while above threshold one sees various regions with net gain. As pumping is further increased, the net gain moves toward higher energies. The shift can be explained by considering two effects that arise due to strong coupling and pumping. 
	
\begin{figure}
		\centering
		\includegraphics[width=0.73\columnwidth]{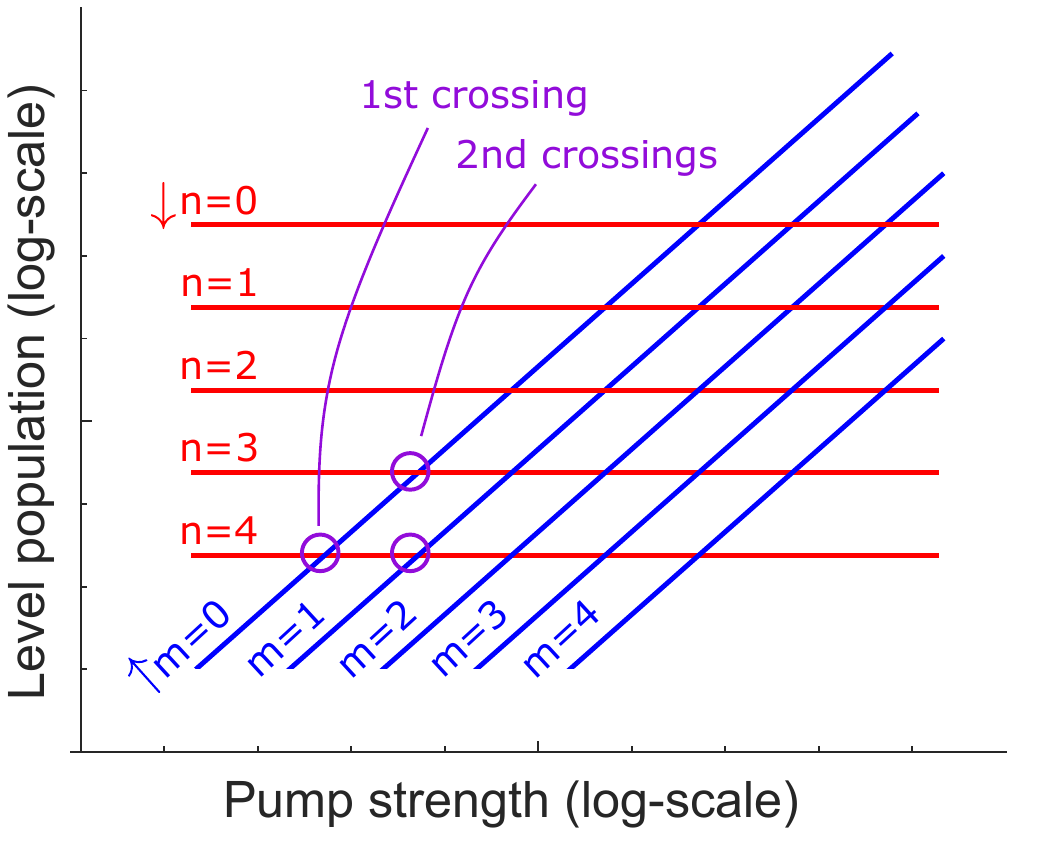}
		\caption{Schematic cartoon of successive inversion of sublevel transitions when vibrational sublevels are thermally populated. Blue and red solid lines represent the $m$ and $n$ levels for transitions $n-m$ from excited state manifold ($m$) to the ground state manifold ($n$) of an organic emitter. As pump strength is increased, the first transition to undergo population inversion is the $4-0$ transition, which corresponds to transitions from the 0th state of the excited state manifold to the 4th state of the ground state manifold.}
		\label{fig:Figsublevels}
	\end{figure}

\begin{figure}
		\centering
		\includegraphics[width=1\columnwidth]{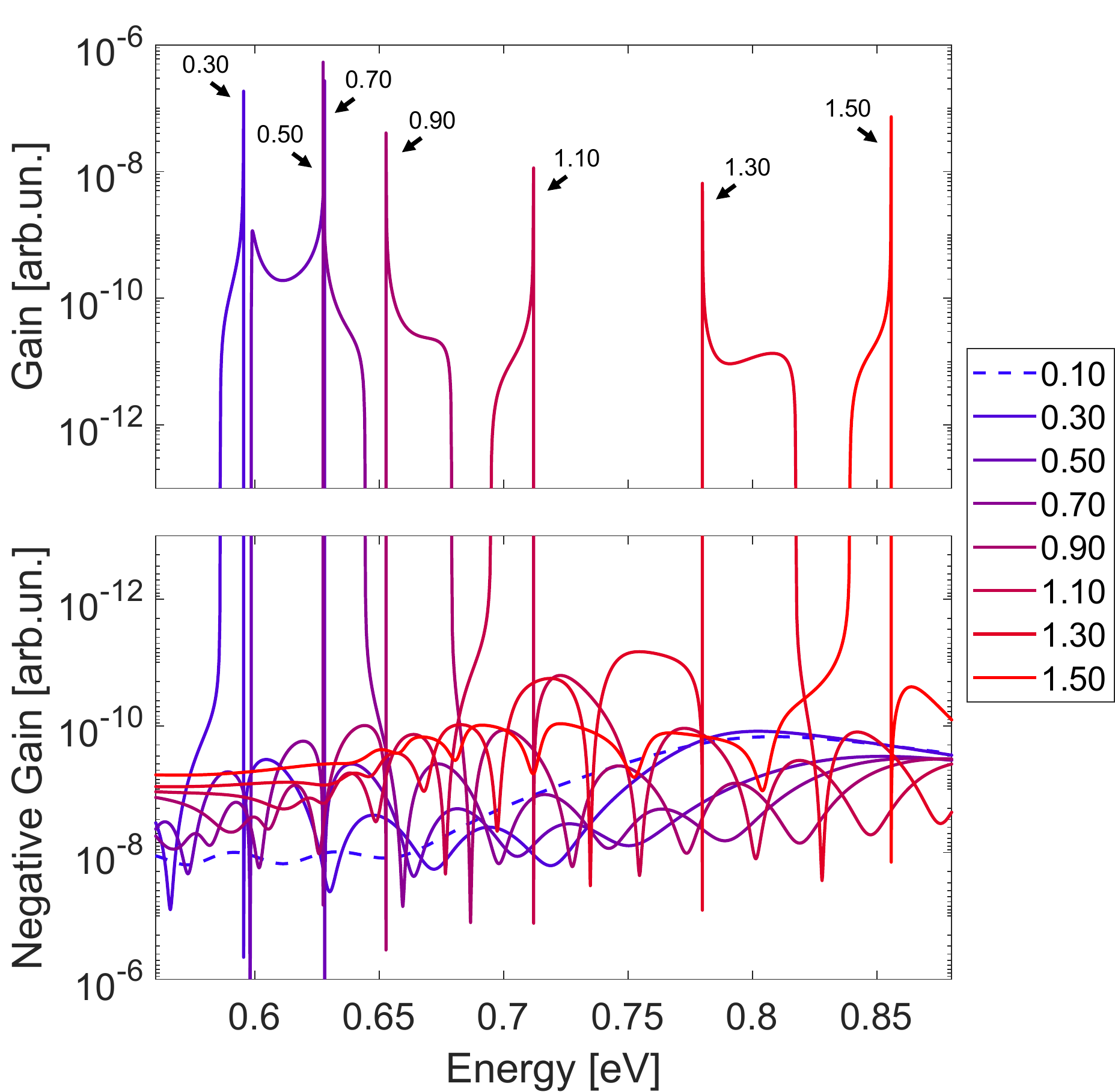}
		\caption{Total gain spectra $G(\nu)$ as a function of pump strength for strong coupling, $\Omega_R=0.4$eV. As in Figs.~\ref{fig:FigXXgain_pump_wc_dense} and~\ref{fig:FigXXgain_pump_wc_sparse}, the two panels show $G(\nu)$ and $-G(\nu)$ on a logarithmic scale. The pump strengths are color coded according to the legend, and all other parameters are as in Table~\ref{tab:params}.}
		\label{fig:FigXXgain_pump_sc_sparse}
	\end{figure}

	First, gain is increased at higher energies as a result of the vibrational sidebands becoming successively inverted with stronger pumping. To understand this we must consider both the relative populations of the ground and excited electronic manifolds, and the populations of the vibrational sublevels labelled by $n$ as illustrated in  Fig.~\ref{fig:Figsublevels}.
	Considering the ground state electronic manifold, we may state the following about the populations of the sublevels.  Unless pumping becomes very strong, the vibrational sublevels will be populated thermally (according to a Boltzmann distribution), so the $n=0$ sublevel is most populated and the higher $n$ are less populated. 
	Turning now to the population of the electronic excited states, we may note that
	the population of the excited state manifold is zero at weak pumping and grows with pumping.
	
	Therefore, with increasing pumping the population of the excited state manifold will start to exceed the populations of the ground-state-manifold sublevels, such that the first sideband to get inverted corresponds to transitions from the 0th state of the excited manifold to the highest $n$ state of the ground state manifold (the ($n-0$) transition), see Fig.~\ref{fig:Figsublevels}. That first inverted transition is also the transition involving the smallest energy difference, thus it leads to gain at lower energies. As the pumping increases further, sidebands with successively smaller $n$ additionally become inverted, thus one starts to get gain at higher energies. At the same time the excited state sublevels with $m>0$ can also become inverted. Both these processes mean ($n-m$) transitions with larger energy difference are inverted, shifting the gain toward higher energy~\cite{Strashko2018}.

\begin{figure*}
		\centering
		\includegraphics[width=0.9\textwidth]{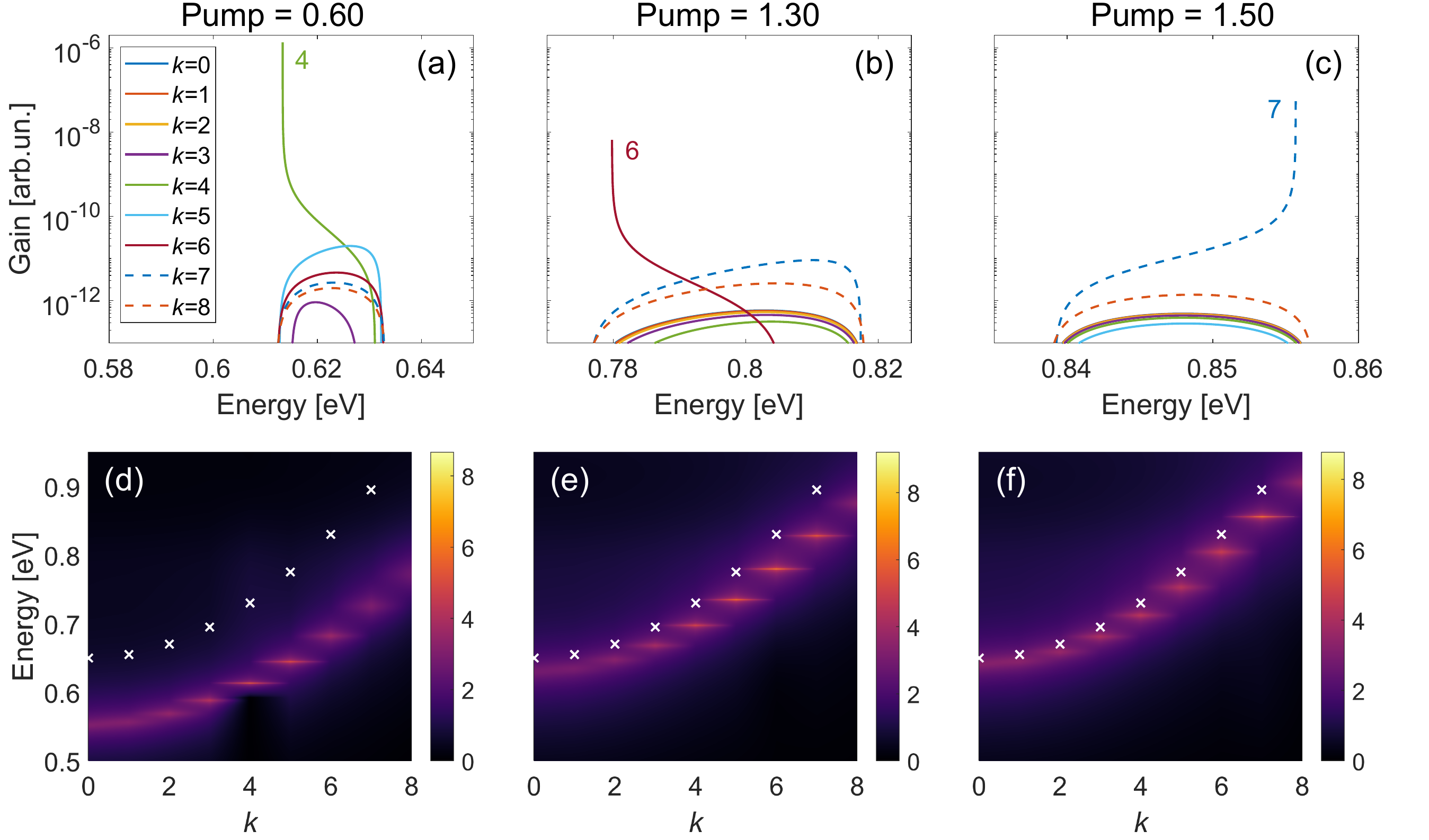}
		\caption{(a-c) Momentum-resolved gain and (d-f) photoluminescence spectra for selected pump strengths for the strong coupling case $\Omega_R=0.4$eV. (a-c) The numbers next to the gain peaks label the $k$ of the mode(s) that are lasing in the steady state. (d-f) The white crosses mark the uncoupled cavity modes. The color scale is logarithmic. The pump strengths are (a,d) $0.60~\Gamma_\uparrow/\Gamma_\downarrow$, (b,e) $1.30~\Gamma_\uparrow/\Gamma_\downarrow$, and (c,f) $1.50~\Gamma_\uparrow/\Gamma_\downarrow$. }
		\label{fig:FigXXpl_gain_pump_sc_kdependent}
	\end{figure*}

	The second effect to be considered is that, not only the gain but also the energies of the strongly-coupled normal modes evolve with pump strength. As shown by the momentum-resolved results in Fig.~\ref{fig:FigXXpl_gain_pump_sc_kdependent}(a-c), for instance when the pump strength is increased from 1.30 to $1.50~\Gamma_\uparrow/\Gamma_\downarrow$, the gain spectrum moves from 0.78--0.82~eV to 0.84--0.86~eV. The photoluminescence spectra in Fig.~\ref{fig:FigXXpl_gain_pump_sc_kdependent}(d-f) show that the dips associated with the polaritonic mode frequencies move to higher energy as the whole dispersion blue shifts upon increasing pump. The blue shift originates from saturation of the two-level systems~\cite{arnardottir_multimode_2020}, reducing the effective coupling strength, ultimately recovering the uncoupled photon mode dispersion at very strong pumping. 
	
	Summarizing, the shifting of the lasing mode toward higher energies upon increasing pumping in the strong coupling case is caused by combined effects of the gain moving toward higher energies, as well as the blueshifting of the polariton mode frequencies.

	\section{Multimode lasing dynamics with pulsed pump}
	\label{sec:pulsed}
	Next we consider the case of pulsed excitation, which is relevant for typical experiments on organic exciton-polariton systems. We excite the system with a 4~ps pulse that has a Gaussian time dependence, and study the dynamics of mode switching as a function of pump strength (amplitude). All other parameters are as in Section~\ref{sec:mmlasing}.
	
	\subsection{Weak coupling regime}
	
\begin{figure*}
		\centering
		\includegraphics[width=1\textwidth]{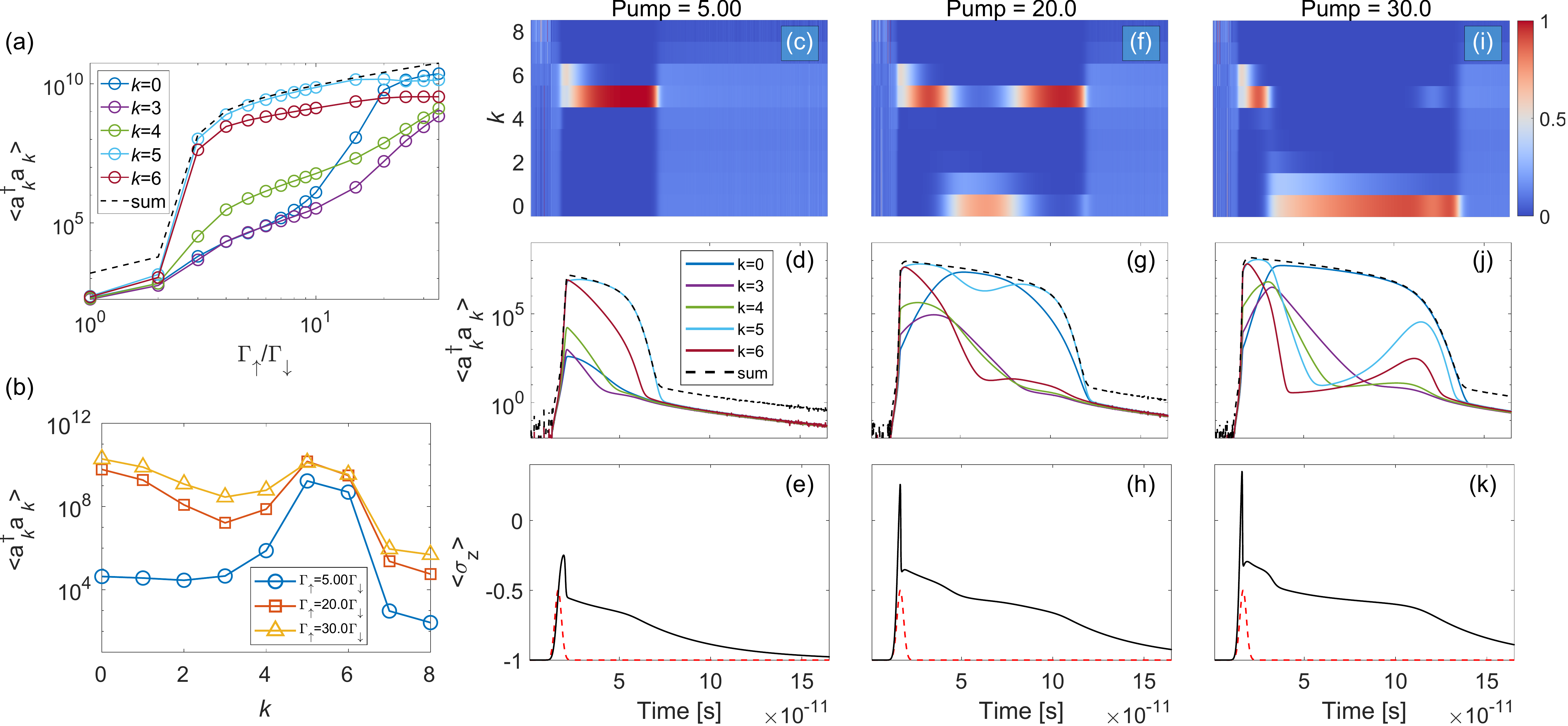}
		\caption{Mode competition with pulsed excitation for weak coupling, $\Omega_R=0.1$eV. (a) Time-integrated pump dependence curves and (b) occupations at different $k$ modes for selected pumps. On the right (c-k): (Top and middle row) time evolution of photon population at different modes and (bottom row) population inversion of the two-level system at various pump strengths. The excitation pulse has a Gaussian time dependence centered at around 16~ps with full width at half maximum of around 4~ps. The pulse profile is shown in (e,h,k) by the red dashed line (in arbitrary units). The peak pump strengths are (c-e) $\Gamma_\uparrow/\Gamma_\downarrow=5.00$, (f-h) $20.0$, and (i-k) $30.0$. All other parameters are provided in Table~\ref{tab:params}. Note that here the time axis is in linear scale.}
		\label{fig:Figpulswc}
	\end{figure*}

	In the weak coupling regime, at low pump strength (Fig.~\ref{fig:Figpulswc}(c-d)), lasing occurs at $k=5$ until decay causes lasing to cease. At higher pump strengths, as shown in Fig.~\ref{fig:Figpulswc}(f-g) and Fig.~\ref{fig:Figpulswc}(i-j), the lasing mode switches from $k=5$ to $k=0$ similar to the continuous pump case discussed in Section~\ref{sec:mmlasing}. However, here the lasing mode switches from $k=0$ back to $k=5$ as the population decays. 
	Throughout the studied pump range, the higher $k=6$ mode gains large population at the onset of lasing. This is also visible in the time-integrated threshold curve and populations in Fig.~\ref{fig:Figpulswc}(a-b): occupation of the $k=6$ mode follows quite closely that of the $k=5$ mode. As shown by Fig.~\ref{fig:Figpulswc}(h,k), at higher pump strengths the two-level systems have very high inversion before the onset of lasing, which is likely connected to excess gain, enough to trigger lasing in several modes.
	
	As noted above, considering time-dependent gain spectra present some issues with interpretation~\cite{Eberly77}.
	The steady-state spectra presented in Section~\ref{sec:gain} can however provide some intuition on how the gain and photoluminescence might shift under time-dependent population inversion. For example, at weak coupling, the peak of the gain switched from $k=5$ to $k=0$ mode when the (continuous) pump strength was increased from 0.30 to 0.70. Here in the pulsed case, the lasing peak switches  in a qualitatively similar way (both as a function of pump strength as well as in time).

	\subsection{Strong coupling regime}
	
\begin{figure*}
		\centering
		\includegraphics[width=1\textwidth]{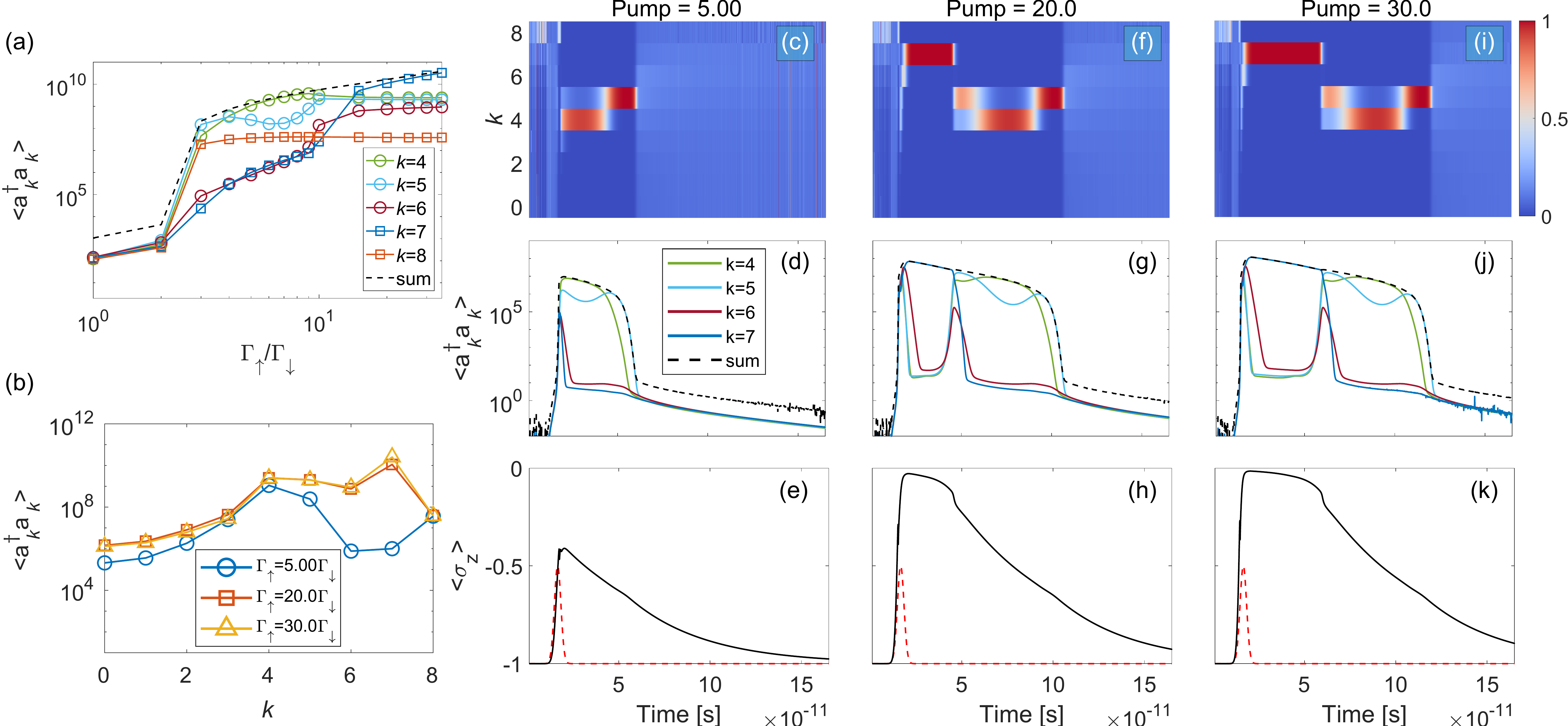}
		\caption{Mode competition with pulsed excitation for strong coupling, $\Omega_R=0.4$eV. (a) Time-integrated pump dependence curves and (b) occupations at different $k$ modes for selected pump powers. On the right (c-k): (Top and middle row) time evolution of photon population at different modes and (bottom row) population inversion of the two-level system at various pump strengths. The excitation pulse has a Gaussian time dependence centered at around 16~ps with full width at half maximum of around 4~ps. The pulse profile is shown in (e,h,k) by the red dashed line (in arbitrary units). The peak pump strengths are (c-e) $\Gamma_\uparrow/\Gamma_\downarrow=5.00$, (f-h) $20.0$, and (i-k) $30.0$. All other parameters are provided in Table~\ref{tab:params}. Note that here the time axis is in linear scale. }
		\label{fig:Figpulssc}
	\end{figure*}

	At strong coupling, for pulsed excitation we observe again rich switching dynamics. At low pump strength, as shown in Fig.~\ref{fig:Figpulssc}(c-d), lasing starts at $k=4$ mode and switches to a higher $k=5$ mode before the population decays. At higher pump strengths, lasing starts in the high $k=7$ mode and undergoes switching via $k=5$ to $k=4$ and back to $k=5$. The duration of the $k=7$ lasing coincides with the time that $\langle\sigma_z\rangle$, shown in Fig.~\ref{fig:Figpulssc}(h,k), remains clamped close to $0$ by polariton lasing.  This duration grows upon increasing the pump strength. The switching sequence to the lower $k$ modes, 5--4--5, occurs during the decay of $\langle\sigma_z\rangle$ in a time scale that we observe to be roughly independent of the pump strength (we confirmed this up to pump amplitude $\Gamma_\uparrow/\Gamma_\downarrow=150$). This time scale is around 40~ps which corresponds to the decay times of spontaneous emission and cavity modes (Table~\ref{tab:params}).
	
	Similar to the weak coupling case, for a short time at the onset of lasing, multiple modes are highly occupied, which contributes to the time-integrated populations shown in Fig.~\ref{fig:Figpulssc}(a-b).

	\section{Discussion}
	\label{sec:disc}
	
	By simulating the time evolution of the photon population, we have explored the dynamics of mode switching, behavior which could not be seen if only the steady-state result is studied. For example, in Fig.~\ref{fig:FigXX2}(c,f), the steady-state result would show single mode ($k=0$) lasing whereas the time evolution reveals that a higher-energy mode ($k=5$) is lasing at earlier times. Furthermore, when comparing simulations to experiments dealing with fast dynamics, the experimental measurement is often restricted to observing the time-integrated spectrum, which may lead to incorrect interpretations of the properties of the system (see Appendix~\ref{app:timeintegrated}). Especially in experiments with fast pulsed excitation, it is common to record time-integrated spectra. Here we have shown that for pulsed excitation the lasing mode can temporally switch such that the time-integrated spectra will exhibit several lasing peaks (or a broadened peak with sub-structure, depending on the mode spacing and resolution of the measurement apparatus). Resolving the time evolution is therefore important in describing the behavior of such systems both in simulations and in experiments.
	
	From the gain calculations we can generally conclude that lasing occurs at a mode (or modes) where there is a sharp positive peak in the gain spectrum. The results suggest that if several modes have positive gain, the mode with the highest positive gain peak will win out the lasing mode competition. In the weak coupling case the highest peak in the gain spectrum simply switches from a higher $k=5$ mode into the $k=0$ mode, given by the different thresholds and slope efficiencies of each mode, In the strongly coupled system, the observed shift of the lasing mode towards higher $k$ modes originates from the shifting of both the gain and the polariton modes toward higher energies as pump is increased. 
	
	We believe that observing mode competition effects such as those shown here is realistic in current experiments. In this paper we used a loss rate of 0.1~meV for both the cavity modes and the excitons. This corresponds to lifetimes of $\sim40$ps (see Table~\ref{tab:params} for all simulation parameters). Under continuous pumping, the lasing starts at around $t=100$~ps ($t=10$~ps) in the weak (strong) coupling case. As shown in Fig.~\ref{fig:FigSXX3}, the time that it takes for the mode switching to occur depends strongly on the pump strength. Nevertheless, we can take examples from our data. In the weak (strong) coupling case with continuous pumping, at pump strength $\Gamma_\uparrow/\Gamma_\downarrow=0.70$ ($1.30$), the switching occurs at around $t=1000$~ps ($t=100$~ps), i.e., after 10 times the time required for lasing to start, or 25 times (2.5 times) the cavity mode lifetime. Another example can be drawn from the pulsed excitation scheme, where the lasing mode is shown to undergo several switching events within 80--120~ps, or 2--3 times the cavity lifetime, from the arrival of the excitation pulse. The temporal switching of lasing modes could be resolved by, e.g., a streak camera.
	
	We note that the cavity lifetime used in the simulations is rather long for an organic polariton system. Cavity lifetimes of the order 10--100~ps are typical for inorganic polariton systems, whereas in organic systems, the lifetimes are typically in the picosecond scale~\cite{keeling_boseeinstein_2020}. The choice of simulation parameters in the present study is based on our earlier works~\cite{Strashko2018,arnardottir_multimode_2020}, and we stress here that the results are similar for shorter lifetimes. 
	
	Multimode polariton lasing could potentially be of use for technological applications. Mode switching phenomena could be used in optical information processing in a similar fashion as polariton switches~\cite{gao_polariton_2012,amo_excitonpolariton_2010} and transistors~\cite{zasedatelev_room-temperature_2019,ballarini_all-optical_2013} (based on on/off-switching of the output intensity of a single-mode condensate) have been proposed. Furthermore, when a system is at a point where multiple modes are lasing in the steady state, even a small disturbance in the operating conditions may push the system from one configuration to another, which could be utilized in sensing applications.
	
	\begin{acknowledgments}
		AJM and PT acknowledge support by the Academy of Finland under project numbers 303351, 307419, 327293, 318987 (QuantERA project RouTe), 318937 (PROFI), and 320167 (Flagship Programme, Photonics Research and Innovation (PREIN)), and by Centre for Quantum Engineering (CQE) at Aalto University. AJM acknowledges financial support by the Jenny and Antti Wihuri Foundation and ETH Zurich Postdoctoral Fellowship. KBA and JK acknowledge financial support from EPSRC program ``Hybrid Polaritonics'' (EP/M025330/1). KBA acknowledges support from The RSE Saltire Research Award.
	\end{acknowledgments}
	
	\newpage 
	
	\appendix

	\section{Details of generalized Gell-Mann matrices}
	\label{app:GGM}
	In this appendix we provide mathematical details of the generalized Gell-Mann matrix representation of the molecules. 
	
	Gell-Mann matrices which have previously been used, e.g., in describing the strong interaction in particle physics~\cite{gell-mann_symmetries_1962,stone2009mathematics}. Their commutation and product rules are given by
	\begin{align}
		\lambda_i=\lambda_i^\dag,\qquad \text{Tr}\lambda_i = 0,\qquad \text{Tr}[\lambda_i\lambda_j] = \delta_{ij}, \nonumber\\
		[\lambda_i, \lambda_j] = 2if_{ijk} \lambda_k,\qquad
		\lambda_i \lambda_j = \zeta_{ijk} \lambda_k + \frac{2}{N_\lambda} \delta_{ij},\nonumber
	\end{align}
	where we may write $\zeta_{ijk}=if_{ijk} + t_{ijk}$ and $N_{\lambda}$ is the dimension of the matrices. The tensors $f_{ijk}$ and $t_{ijk}$ are the symmetric and antisymmetric structure constants, respectively. Gell-Mann matrices allow for expanding any linear $N_{\lambda} \times N_{\lambda}$ operator in this basis as
	\begin{equation}
		A = \frac{1}{N_{\lambda}} \text{Tr}[A]\mathbbm{1}_{N_{\lambda}} + \frac{1}{2}\text{Tr}[A\lambda_i]\lambda_i.
	\end{equation}
	
	Using the above results, we may identify the coefficients in the second-order cumulant equations (Eq.~\eqref{eq:secondcumulants}) as follows:
	\begin{align}
		&\phi_i = \frac{2i}{\Nv}f_{ipr} \sum_\mu \gamma_p^\mu \gamma_r^{\mu \ast}, \nonumber \\ 
		&\beta_{ij} = \frac{1}{2}B_p(f_{i_x pj}-i f_{i_y pj}), \nonumber \\  
		&X_{ij} = \xi_{i_xj_x}+i\xi_{i_yj_x},  \nonumber \\
		&\xi_{ij} = 2 f_{ipj} A_p + i \left(f_{ips}\zeta_{rsj} + f_{ris}\zeta_{spj}\right)\sum_\mu \gamma_p^\mu \gamma_r^{\mu \ast}. \nonumber
	\end{align}

	\section{Time-integrated threshold curves}
	\label{app:timeintegrated}
\begin{figure*}[b]
		\centering
		\includegraphics[width=0.7\textwidth]{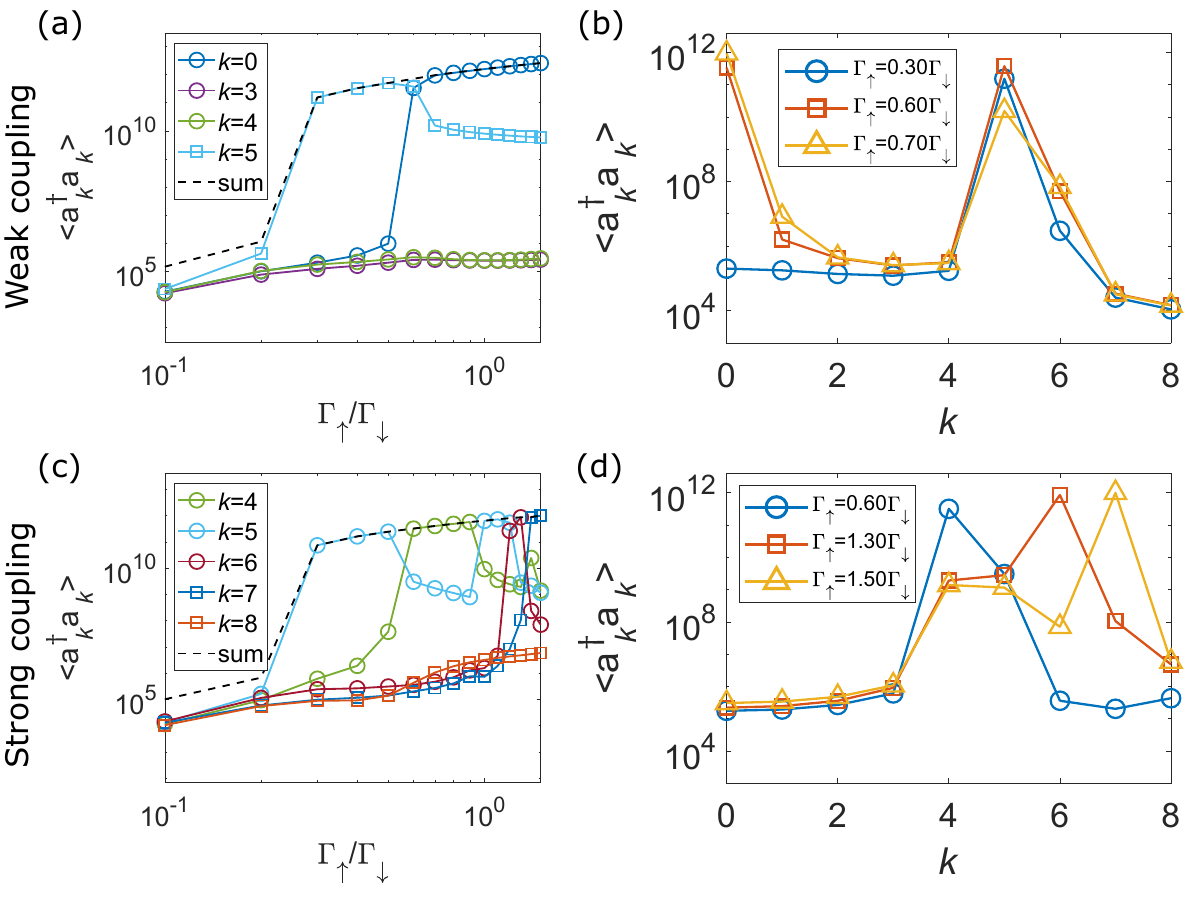}
		\caption{Time-integrated results with continuous excitation for weak and strong coupling. (a,c) Pump dependence curves and (b,d) occupations at different $k$ modes. The mode occupations in (b,d) are shown for selected pump powers. The weak coupling case ($\Omega_R=0.1$eV) is shown on the top row and the strong coupling case ($\Omega_R=0.4$eV) on the bottom row. }
		\label{fig:FigSXX5}
	\end{figure*}

	In this appendix we present threshold curves and population distributions obtained by time-integrating the photon mode populations until steady state has been reached. The results shown in Fig.~\ref{fig:FigSXX5} represent a common case in experimental measurements of a system with fast dynamics, where, instead of having access to the temporal evolution of the photon populations, mode populations are collected by integrating over a finite time. Our results show that observing multiple peaks in the time-integrated mode populations can refer to temporal switching between modes. 
	
	\bibliographystyle{apsrev4-1_abbrv}
	\bibliography{bibfile}

\end{document}